
\parindent = 20 pt
\hsize = 6.5 in
\baselineskip = 12 pt

\centerline{NONRADIAL SOLUTIONS OF A SEMILINEAR}
\centerline{ELLIPTIC EQUATION IN TWO DIMENSIONS}
\vskip .1 in
\centerline{Joseph Iaia and Henry Warchall}
\centerline{Department of Mathematics}
\centerline{University of North Texas}
\centerline{Denton, Texas, 76203-5116}
\vskip .2 in

\hangindent = .75 in
\hangafter = 1
\noindent
Abstract : We establish existence of an infinite family of
exponentially-decaying non-radial  $C^2$
solutions to the equation  $\Delta u+f(u)=0$  on  $R^2$ for a large class
of nonlinearities $f$. These solutions have the form
$u(r,\theta )=e^{im\theta }w(r)$,  where  $r$  and  $\theta$  are polar
coordinates,  $m$  is an integer, and  $w:[0,\infty )\to R$  is
exponentially decreasing far from the origin. We prove there is a solution
with each prescribed number of nodes.
\baselineskip = 17 pt
\hsize = 6.5 in
\vsize = 8.5 in
\voffset = 0 in
\hoffset = 0 in
\vskip .2 in
\centerline{ 1.  INTRODUCTION}
\vskip .2 in
        We consider the semilinear elliptic equation  $\Delta u+f(u)=0$,
where the nonlinearity $f:C\to C$  has the property that
$f(se^{i\psi })=f(s)e^{i\psi }$  for all real  $s$  and  $\psi$.
The behavior of such a function is determined by its restriction to real
arguments, and henceforth we refer only to the restriction of  $f$
to the real axis, which is necessarily odd.  We look for
$C^2$  solutions  $u:R^N\to C$  such that  $u(x)\to 0$  as
$\left| x \right|\to \infty $.  Interest in these
solutions stems from their role as the spatial profiles of localized
standing-wave solutions to nonlinear evolution equations, including the
nonlinear Schr\"odinger and nonlinear wave equations. The set of
spherically-symmetric (``radial'') solutions has been extensively studied
(see [1] - [4], [6] - [12]).

        In this paper we use ordinary-differential-equation arguments to
establish the existence of solutions that do not have rotational symmetry,
in the case of  $N=2$ spatial dimensions.  We make use of an ansatz due
to P.-L. Lions ([8]) to reduce the study of the partial differential
equation to that of an ordinary differential equation.  Specifically, we look
for solutions  $u$  of the form  $u(r,\theta )=e^{im\theta }w(r)$,  where
$r$  and  $\theta$ are polar coordinates for $R^2$,  $m$  is a nonzero
integer, and  $w:[0,\infty )\to R$.  Substituting this ansatz into the
elliptic equation for  $u$  yields the ordinary differential equation
$w''+{\textstyle{1 \over r}}w'-{\textstyle{{m^2} \over {r^2}}}w+f(w)=0$
for  $w$.   Without loss of generality, we henceforth assume that  $m$  is
a positive integer.

        A method based on variational arguments for proving existence
of solutions obeying this ansatz (and higher-dimensional generalizations) is
outlined in [8].  In [7] solutions are explicitly computed for a
piecewise-linear nonlinearity  $f$.  Here we use shooting arguments that
parallel those in [10] to directly establish the existence of $C^2$ solutions
for a large class of nonlinearities. Our assumptions on $f$ are essentially
the same as those in [10].  We suppose that the restriction of  $f$  to
real arguments is an odd locally Lipschitz-continuous function with
$-\infty <- \sigma^{2} \equiv \mathop {\lim}\limits_{s\to 0}{{f(s)} \over s}
\leq 0$, and in case $\sigma = 0$ we require that $f(s) < 0$ for small
positive $s$.  We assume that the primitive $F(s)\equiv \int_0^s {f(t)\,\,dt}$
has exactly one positive zero $\gamma$, and that $f(s)> 0$ for
$s \in [\gamma, \infty).$ We also assume that
$f(s)=\kappa \left| s \right|^{p-1}s+g(s)$,  where  $\kappa$  is a positive
constant, $p>1$,  and  $s^{- p}g(s)\to 0$  as  $s\to \infty $.   The results
in [8] are based on the hypothesis  $\sigma =0$, which results in algebraic
decay of solutions
\vfill\eject
\noindent
far from the origin, whereas our solutions for $-\sigma^{2} <0$  have
exponential decay. We make the assumption  $- \infty < -\sigma^{2} \equiv
\lim_{s \to 0} {f(s) \over s} \leq 0$ for simplicity. The conclusion of the
Main Theorem which follows remains true if that assumption is replaced by
the requirement that ${f(s) \over s}$ is bounded below, with $f$ negative
for small positive $s$.

        Because of the rather strong singularity at  $r=0$  in the
ordinary differential equation for $w$,  it is not immediately apparent
how to formulate a well-posed initial value problem for  $w$
with initial conditions given at  $r=0$.   To gain insight, we make the
change of variable $w(r)=r^mv(r)$  to obtain the equation
$$v''+{\textstyle{{2m+1} \over r}}v'+
  {\textstyle{1 \over {r^m}}}f(r^mv)=0 \eqno (1.1)$$
for  $v$,  which, by virtue of the condition on  $f$  at zero, has well-posed
initial value problems obtained by specifying
$$v(0)=d, \ {\rm and} \ v'(0)=0   \eqno (1.2) $$
(see Section 2).   If  $v$  is a $C^2$ solution of such an initial value
problem, it follows that $w(r)=r^m v(r)$  is a $C^2$ solution of the initial
value problem
$$w''+{\textstyle{1 \over r}}w'-{\textstyle{{m^2} \over {r^2}}}w+f(w)=0
        \eqno (1.3) $$
subject to
$$\mathop {\lim }\limits_{r\to 0^+}{\textstyle{1 \over {r^m}}}w(r)=d
 \ {\rm and } \
\mathop {\lim }\limits_{r\to 0^+}{\textstyle{1 \over {r^{m-1}}}}w'(r)=md,
  \eqno (1.4) $$
and that the corresponding function  $u(r,\theta )=e^{im\theta }w(r)$  is
$C^2$ on  $R^2$.

        Note that, although the initial value problem (1.1)-(1.2) is
superficially similar to the much-studied radial problem consisting
of the differential equation
$$v''+{\textstyle{{N-1} \over r}}v'+f(v)=0   \eqno (1.5)$$
subject to initial conditions (1.2), there is a significant difference between
the terms  ${\textstyle{1 \over {r^m}}}f(r^mv)$  and  $f(v)$.  Interpreting
the differential equations as equations of motion for a point with position
$v(r)$  at time  $r$,  we note that according to (1.2) the system is
released from rest with initial displacement  $d$.   The system moving under
(1.1) initially experiences the repulsive force
$\mathop {\lim }\limits_{r\to 0^+}{\textstyle{{-1} \over
{r^m}}}f(r^mv)= \sigma^{2} d $  determined by the behavior of
$f$  at the origin, whereas the system moving under (1.5) initially experiences
the force $-f(d)$,  which is attractive for the values of  $d$  that yield
solutions that decay at infinity.  The character of the problem
(1.1)-(1.2) is thus quite different from that of (1.5) with (1.2), and
necessitates a separate analysis.
\vskip .2 in
        We prove the following main theorem:
\vskip .1 in
\noindent
{\bf MAIN THEOREM}:  Let the nonlinearity $f$ have the properties specified.
Then, for each nonnegative integer  $n$,  there
is a positive number $d $  and a $C^2$ solution  $w$  to  (1.3)-(1.4)  such
that $\mathop {\lim }\limits_{r\to \infty }w(r)=0$ and  $w$  has exactly
$n$  positive zeros.
\vskip .2 in
   In section 2, we show that the initial value problem (1.3) - (1.4) has
solutions for all $r > 0$. In section 3, we prove that there are values of
$d$ for which (1.3) - (1.4) has solutions that are positive for all $r> 0$.
In section 4, we show that, similarly, there are values of $d$ for which
(1.3) - (1.4) has solutions with any prescribed number of zeros. Section 5
contains the proof of the main theorem. Finally, in section 6, we show that
our solutions decay exponentially far from the origin if $-\sigma^{2} < 0.$

   In the following we denote by $\alpha$ the smallest positive zero of $f$,
and by $\beta$ the largest positive zero of $f$.  We thus have
$0 < \alpha \leq \beta < \gamma.$ Also,  we write $r \to 0$ and $d \to 0$
to mean $r \to 0^{+}$ and $d \to 0^{+},$ respectively.

   We make repeated use of $Pohozaev's$ $Identity$ :
$$  r^{2}( {1 \over 2} w'^{2} + F(w)) |_{r_{1}}^{r_{2}}  =
  {m^{2} \over 2} w^{2} |_{r_{1}}^{r_{2}}    +
      2 \int_{r_{1}}^{r_{2}} sF(w(s)) \, ds $$
which results from multiplying (1.3) by $r^{2}w'$ to obtain
$  ({1 \over 2} w'^{2})' - ({m^{2} \over 2} w^{2})' + r^{2}( F(w))' = 0, $
and then integrating on $(r_{1}, r_{2})$.

\vskip .2 in
\centerline{2. GLOBAL EXISTENCE OF SOLUTIONS TO THE INITIAL VALUE PROBLEM}
\vskip .2 in

We first observe that with the relationship
$$    w(r)=r^mv(r),         \eqno (2.1)  $$
which will be used throughout the paper, the initial value problems (1.1)-(1.2)
and (1.3)-(1.4) are
equivalent.  To see this, note first that a simple calculation shows that
if  $v$ satisfies (1.1)-(1.2) then  $w$  satisfies (1.3)-(1.4).
Conversely, it is easy to see that if  $w$  satisfies (1.3)-(1.4) then
$v$  satisfies (1.1) and  $v(0)=\mathop {\lim }\limits_{r\to 0}v(r)=d$.
To show  $v'(0)=0$ requires a bit more work:

\parindent = 20 pt
\hsize = 6.5 in
\vsize = 8.5 in
\voffset = 0 in
\hoffset = 0 in

Rewriting (1.3) gives
$$(r^{2m+1}({w \over r^{m}})')' = -r^{m+1}f(w).$$
Integrating on $(0,r)$ and noting from (1.4) that
$$\lim_{r \to 0} r^{2m+1} ({w \over r^{m}})' =
 \lim_{r \to 0} r^{m+1} w' - m r^{m} w = 0$$ gives
$$ ({w \over r^{m}})' = {-1 \over r^{2m+1}} \int_{0}^{r} s^{m+1} f(w)\, ds.$$
Thus, $$\lim_{r \to 0} v'(r) =
\lim_{r \to 0} { -\int_{0}^{r} s^{m+1} f(w) \, ds \over r^{2m+1}} =
 \lim_{r \to 0} - { f(w) \over w}{ w \over r^{m}} { r \over 2m+1}
= --\sigma^{2} \, \cdot d \, \cdot 0 = 0.$$

\vskip .2 in
To show the small $r$ existence of solutions to (1.1) - (1.2), note that
solutions are fixed points of the mapping $G$ defined by
$$ G(v(r)) = d
 - \int_{0}^{r}{1 \over s^{2m+1}}\int_{0}^{s}t^{m+1}f(t^{m}v(t)) \, dt \, ds.
$$
We will now show that $G$ is a contraction mapping for small
$r$. Suppose $y$ and $v$ are continuous on $[0, T]$ with $ |y|, |v| < C$.
Then
$$ |G(y) - G(v)| \leq
 \int_{0}^{T}{1 \over s^{2m+1}}\int_{0}^{s}t^{m+1}|f(t^{m}y(t)) - f(t^{m}v(t))|
\, dt \, ds$$
$$ \leq  K |y - v|
 \int_{0}^{T}{1 \over s^{2m+1}}\int_{0}^{s}t^{m+1}t^{m} \, dt \, ds
\leq { T^{2} K  \over 4(m+1)} |y-v| $$ where $K$ is the Lipschitz constant
for $f$ on $[-C, C].$
Thus, for $T$ small enough we obtain
$$ |G(y) - G(v)| < \epsilon |y-v| $$ where $\epsilon < 1.$
Thus, by the contraction mapping principle, $G$ has a unique fixed point
for $T$ small, and therefore there exists a solution of (1.1) - (1.2) for
$T$ small and hence of (1.3) - (1.4) for $T$ small.
\vskip .2 in
To show that the solutions to (1.3) - (1.4) exist for all $r > 0$, we show
that $w$ and $w'$ remain finite by considering the quantity
$$ \tilde E(r) =
{ {1 \over 2} w'^{2}(r) + F(w(r)) - F_{0} \over r^{2m-2}}
 + {m^{2} \over 2} {w^{2}(r) \over r^{2m}} $$
where $F_{0} = \min F < 0.$ Here
$$\tilde E(0) =  + \infty  \  {\rm and} \  \tilde E(r) \geq 0. $$
A computation shows
$$ \tilde E' = -{m \over r^{2m-1}} (w' - {mw \over r})^{2}
               - { 2(m-1) [F(w(r)) - F_{0}] \over r^{2m-1} } \leq 0.$$
For each solution $w(r)$, $\tilde E(r)$ is thus a nonnegative, nonincreasing
quantity. Since the term $ F(w(r)) - F_{0}$ is nonnegative, we have that
$$  {w'^{2} \over r^{2m-2}} + { w^{2} \over r^{2m}} \leq 2 \tilde E(r_{0})$$
for all $r \geq r_{0}$ where $r_{0}$ is any positive number. From this and
the small $r$ existence of solutions, it follows that the solution
$w(r)$ exists for all $r> 0.$

\vskip .2 in
\centerline{3. EXISTENCE OF POSITIVE SOLUTIONS }
\vskip .2 in
\noindent
{\bf LEMMA 3.1} : Let $w$ be a nontrivial solution of (1.3) and suppose
$w(r_{0}) = 0.$ Then there exists a smallest $b > r_{0}$ such that
$|w(b)| = \alpha.$ Furthermore, $w \neq 0$ and $w' \neq 0$ on $(r_{0}, b]$.
\vskip  .2 in
\noindent
{\bf PROOF OF LEMMA 3.1} :
A nontrivial solution $w$ cannot vanish on any nonempty open interval,
by uniqueness of solutions to initial value problems.
So, there is an interval $(r_{0}, r_{0} + \epsilon)$ on which either $w>0$
and $w' >0$ or $w<0$ and $w' <0$. Without loss of generality, assume $w>0$
and $w'>0$ on $(r_{0}, r_{0} + \epsilon).$ There are now two possibilities :
either
$$ w {\rm \ has \ a \ positive \ local \ maximum \ at
\ some \ smallest \ value \ of \ } r, {\rm \ say} \ r_{1},  \eqno (3.1) $$
or
$$ w' \geq 0 \ {\rm for \ all} \ r \geq r_{0}. \eqno (3.2)  $$
\vskip .2 in
\noindent
If (3.1) holds, then at $r_{1}$ we have $w'(r_{1}) = 0$ and $w''(r_{1}) \leq
0$.
Substituting into (1.3) gives
$$ - {m^{2} w(r_{1}) \over r_{1}^{2}} + f(w(r_{1})) \geq 0. $$
Thus $ f(w(r_{1})) \geq {m^{2} w(r_{1}) \over r_{1}^{2}} > 0$ which
implies $w(r_{1}) \geq \alpha.$ Therefore, there exists a smallest
$b > r_{0},$ with $b <r_{1},$ for which $w(b) = \alpha$.
\vskip .2 in
\noindent
If (3.2) holds then there are two possibilities.
Either there exists a smallest $b > r_{0}$ such that $w(b) = \alpha$ or
$$ w'(r) \geq 0, \ {\rm and } \ 0 < w(r) < \alpha \ {\rm for \ all} \ r >r_{0}.
\eqno (3.3) $$
We claim that (3.3) is impossible. If indeed (3.3) holds,
then $f(w) < 0$ for all $r > r_{0}$ and so we have
$$ w'' + {w' \over r} - {m^{2} w \over r^{2}} = - f(w) > 0$$
or equivalently, $$ (r^{2m+1}v')' \geq 0$$ for all $r > r_{0}$. Integrating
twice on $(r_{0},r)$ gives :
$$ v(r) > v(r_{0}) + {r_{0} \over 2m}v'(r_{0})[1 - ({r_{0} \over r})^{2m}]. $$
If $r_{0} =0$ then $v(r_{0}) = d > 0$ and $v'(r_{0}) = 0$, and so we have
$$ w(r) \geq dr^{m}.$$ If $r_{0} > 0$ then $v(r_{0}) = 0$ and $v'(r_{0}) > 0$,
so we have
$$  w(r) > {r_{0} \over 2m}v'(r_{0})[r^{m} - {r_{0}^{2m} \over r^{m}}].$$
In either case, $w$ grows without bound as $r$ increases, contradicting (3.3).

   We have thus established that there exists a smallest $b > r_{0}$ such that
$w(b) = \alpha$. It is clear that $w(r) >0$ and $w'(r) \geq 0$ for all
$r$ in $(r_{0}, b]$.

   To show that $w' \neq 0$ on $(r_{0}, b]$, suppose on the contrary that
$w'(r_{2}) = 0$ for some $r_{2}$ in $(r_{0}, b].$ Then (1.3) gives
$ w''(r_{2}) = {m^{2} \over r_{2}^{2}} - f(w(r_{2})) > 0.$ Thus, $w(r_{2})$
is a local minimum. This contradicts the fact that $w(r) > 0$ and $w'(r) \geq
0$
for all $r$ in $(r_{0}, b].$
\vskip .2 in
\noindent
This completes the proof of Lemma 3.1.
\vskip .4 in
\noindent
{\bf LEMMA 3.2} : Let $w$ be the solution to (1.3)-(1.4) for $d>0$. Let
$b_{d}$ be the smallest positive value of $r$ for which $w(r) = \alpha.$
Then as $d \rightarrow 0$, $b_{d} \rightarrow \infty.$
\vskip .2 in
\noindent
{\bf PROOF OF LEMMA 3.2} :
Because $f$ is bounded below and $\lim_{w \to 0} {f(w) \over w} = -\sigma^{2} <
0$,
there exists $M > 0$ such that
$$ { f(w) \over w} \geq -M \ {\rm for \ all} \ w.$$
Therefore,
$$ w'' + {w' \over r} - {m^{2} w \over r^{2}} = - f(w) \leq M w
     \ \ {\rm for} \ 0 \leq w \leq \alpha $$
or equivalently
$$ v'' + {2m+1 \over r} v' \leq M v  \ {\rm for} \ 0 \leq r \leq b_{d}.
\eqno (3.5)$$
Since $v > 0$ on $[0, b_{d}]$, dividing by $v$ gives
$$ {v''\over v} + {2m+1 \over r} {v' \over v} \leq M.$$
Letting $$y = \log v  \eqno (3.6)$$ we obtain
$$ y'' + {2m+1 \over r} y' \leq y'' + y'^{2} + {2m+1 \over r} y' \leq M.$$
Thus,
$$ (r^{2m+1}y')' \leq M r^{2m+1}.$$
Integrating on $(0, r)$ and noting that
$\lim_{r \to 0} r^{2m+1} y' = 0$ gives
$$ r^{2m+1} y' \leq  {M \over 2(m+1)} r^{2m+2} .$$
Integrating again on $(0, r)$ gives
$$  \log {v \over d} \leq {M \over 4(m+1)} r^{2}.$$
Thus,
$$  w \leq dr^{m} e^{{M \over 4(m+1)} r^{2}}
\ {\rm for} \ 0 \leq r \leq b_{d}.   \eqno (3.7)$$
Evaluating at $r = b_{d}$ gives
$$ \alpha \leq db_{d}^{m} e^{{M \over 4(m+1)} b_{d}^{2}}. $$
Thus, as $d \to 0$, we have $b_{d} \to \infty.$
\vskip .2 in
\noindent
{\bf LEMMA 3.3} : Let $w$ and $b_{d}$ be as in Lemma 3.2. Let $a_{d}$ be the
smallest
positive number such that $w({a_{d}}) = { \alpha \over 2}.$ Then
$ b_{d} - a_{d} \geq K^{2} > 0$ where $K^{2}$ is a constant that
is independent of $d$ if $d$ is small.
\vskip .2 in
\noindent
{\bf PROOF OF LEMMA 3.3} :
It follows by evaluating (3.7) at $r=a_{d}$
that $\lim_{d \to 0} a_{d} = \infty.$
Pohozaev's Identity for $(0, r)$ is
$$ {1 \over 2}r^{2} w'^{2} + r^{2} F(w) = {m^{2} \over 2} w^{2} +
          2 \int_{0}^{r} sF(w) \, ds. \eqno (3.8)$$
For $a_{d} \leq r \leq b_{d}$ we have $ {\alpha \over 2} \leq w \leq \alpha$
and also $F(w) < 0.$ Thus,
$$ {1 \over 2} w'^{2} + F(w) < {m^{2} \over 2} {\alpha^{2} \over a_{d}^{2}}
     \  {\rm for} \ a_{d} \leq r \leq b_{d}.$$
Now let $C(d) \equiv {m^{2} \over 2} {\alpha^{2} \over a_{d}^{2}} $ and note
$C(d) \to 0$ as $ d \to 0.$
Thus,
$$ {w'^{2} \over C(d) - F(w)} < 2 \ {\rm for} \ a_{d} \leq r \leq b_{d}.$$
We saw in the proof of Lemma 3.1 that $w' \geq 0$ for $0 \leq r \leq b_{d}.$
So, taking square roots of the above and integrating on $(a_{d}, b_{d})$
gives :
$$ \int_{{\alpha \over 2}}^{\alpha}  {ds \over \sqrt{C(d) - F(s)}} =
\int_{a_{d}}^{b_{d}}{ w' \over \sqrt{C(d) - F(w)}} \, dr <  \sqrt{2}(b_{d} -
a_{d}).$$
Now as $d \to 0$ the left hand side of the above approaches the
constant $$   \int_{{\alpha \over 2}}^{\alpha}  {ds \over \sqrt{- F(s)}}   > 0.
$$
Thus for $d$ small enough we have
$$   b_{d} - a_{d} \geq K^{2} > 0.$$ This completes the proof of Lemma 3.3.
\vskip .2 in
\noindent
{\bf LEMMA 3.4} : Let $w$ be the solution of (1.3)-(1.4) for $d> 0.$
For $d$ chosen small enough, we have $0 < w(r) < \gamma$
for all $r >0.$
\vskip .2 in
\noindent
{\bf PROOF OF LEMMA 3.4} : Recall that $w(r) > 0$ and $w'(r) > 0$ for $r$
small. We first claim that if $w(r) < \gamma$ for $r$ in $(0, c)$
then $w(r) \neq 0 $ for $r$ in $(0, c).$

  To establish this fact, suppose that $w(r)=0$ for some $r$ in $(0,c)$ and
let $z_{d} \in (0, c)$ be the smallest such value of $r$. Pohozaev's Identity
on $(0, z_{d})$ is
$$   {1 \over 2} z_{d}^{2} w'^{2}(z_{d}) = 2 \int_{0}^{z_{d}} rF(w(r)) \, dr.$$
Since $0 < w(r) < \gamma$ for $r$ in $(0, z_{d})$, $F(w(r)) < 0$ for
$r$ in $(0,z_{d})$. Thus the right hand side is negative, whereas the left hand
side is nonnegative. Thus, there is no zero of $w(r)$ in the interval
$(0,c)$ if $w(r) < \gamma$ on $(0, c).$

  We next claim that for sufficiently small $d$, $w(r) < \gamma$ for all
$r > 0.$  To establish this fact, suppose there is a smallest value
$c_{d}$ of $r$ such that $w(c_{d}) = \gamma.$
Then, as we just established, $0 < w < \gamma \ {\rm on} \ (0, c_{d}).$
\vskip .1 in
Pohozaev's Identity on $(0, c_{d})$ is :
$$ 0 \leq  {1 \over 2} c_{d}^{2} w'^{2}(c_{d}) = {m^{2} \over 2} \gamma^{2}
    + 2 \int_{0}^{c_{d}} rF(w) \, dr. \eqno (3.9)$$
We will now show that the right hand side is negative for sufficiently small
$d$, whereas the left hand side is nonnegative, resulting in a contradiction
to the supposition $w(c_{d}) = \gamma.$
We have the following inequalities $$ 0 < a_{d} < b_{d} < c_{d}$$ and
$F(w) \leq 0$ on $(0, c_{d}).$
Further, for $ a_{d} \leq r \leq b_{d}$ we have $ {\alpha \over 2} \leq w \leq
\alpha$
and thus $F(w) \leq F({\alpha \over 2}) < 0$ since $F$ is decreasing on
$[ {\alpha \over 2}, \alpha].$
Thus,
$$ \int_{0}^{c_{d}} rF(w) \, dr \leq  \int_{a_{d}}^{b_{d}} rF(w) \, dr
  \leq {1 \over 2} F({\alpha \over 2}) (b_{d}^{2} - a_{d}^{2}). \eqno (3.10)$$

Now from Lemmas 3.2 and 3.3 we have
$$ b_{d}^{2} - a_{d}^{2} = (b_{d} - a_{d})(b_{d} + a_{d})
    \geq K^{2}(b_{d} + a_{d}) \geq K^{2} b_{d} \to \infty \ {\rm as} \ d \to
0.$$
Since $F({\alpha \over 2}) < 0$ we have
$$ \int_{0}^{c_{d}} rF(w) \, dr \to - \infty \ {\rm as} \ d \to 0.$$
Hence, the right hand side of (3.9) is negative for $d$ small enough and
this gives the desired contradiction.

\parindent = 20 pt
\hsize = 6.5 in
\vsize = 8.5 in
\voffset = 0 in
\hoffset = 0 in
\vskip .4 in
\centerline{4. SOLUTIONS WITH PRESCRIBED NUMBER OF ZEROS}
\vskip .4 in

 In this section, we show that there are solutions of (1.3) - (1.4) with an
arbitrarily large number of zeros. To do this, we study the behavior of
solutions as $d$ grows large. In this section, given $\lambda > 0$,
let $z_{\lambda}(r)$ be the solution
of (1.3) - (1.4) with $d \equiv \lambda^{ {2 \over p-1} + m}.$
Then define
$$ y_{\lambda}(r) = \lambda^{-{2 \over p-1}}r^{-m}z_{\lambda}
   ({r \over \lambda}).   \eqno (4.1)  $$
Then $y_{\lambda}$ satisfies
$$ y_{\lambda}'' + {2m+1 \over r} y_{\lambda}' +  \lambda^{-{2 \over p-1} -2}
  r^{-m} f(\lambda^{2 \over p-1} r^{m} y_{\lambda}) = 0   \eqno (4.2) $$
and
$$  y_{\lambda} (0) = 1, \ y_{\lambda}'(0) = 0.   \eqno (4.3) $$
\vskip .2 in
\noindent
Now we make use of the hypothesis
$$ f(w) = \kappa |w|^{p-1}w + g(w)  \eqno (4.4) $$
where $\lim_{|w| \to \infty} { g(|w|) \over |w|^{p}} = 0.$
\vskip .2 in
\noindent
{\bf LEMMA 4.1} : As $\lambda \to \infty$, $y_{\lambda} \to y$ uniformly on
compact subsets of $[0, \infty),$ where $y$ is the solution of
$$y'' + {2m +1 \over r}y'  + \kappa r^{m(p-1)} |y|^{p-1}y = 0 \eqno (4.5) $$
and
$$  y(0) = 1,
 \ y'(0) = 0.       \eqno (4.6)$$
\vskip .2 in
\noindent
{\bf PROOF OF LEMMA 4.1} :
As in Section 2, we can show that for each $\lambda > 0$,
(4.2) - (4.3) has a solution for all $r > 0.$ We
can also define a decreasing energy by
$$ E_{\lambda}(r) = {1 \over 2} (ry_{\lambda}' + my_{\lambda})^{2}
     + {m^{2} \over 2} y_{\lambda}^{2} +
 \lambda^{-{4 \over p-1} - 2}
 {F(\lambda^{2\over p-1} r^{m} y_{\lambda}) \over r^{2m-2}}
 +  2(m-1) \lambda^{- {4 \over p-1} - 2}  \int_{0}^{r}
 {F(\lambda^{2 \over p-1} s^{m}y_{\lambda}) \over s^{2m-1}}  \, ds \eqno (4.7)
$$
Here,
$$ E_{\lambda}' = -ry_{\lambda}'^{2} \leq 0,
   \ {\rm and} \  E_{\lambda}(0) =  \lim_{r \to 0} E_{\lambda}(r) = m^{2}.$$
\vskip .2 in
Now we will show that the integral term in the above energy is bounded below
as $\lambda \to \infty$. This will allow us to bound the other terms from
above and then appeal to the Arzela-Ascoli Theorem to obtain
a convergent subsequence of the $\{ y_{\lambda} \} $.
\vskip .2 in
To this end recall the following
inequalities which were established in Lemma 3.1 and equation (3.6):
$$  dr^{m} \leq w(r) \leq dr^{m}e^{{M \over 4(m+1)}r^{2}}
    \ {\rm for} \ 0 \leq r \leq b_{d} \eqno (4.8) $$
where $w$ is a solution of (1.3)-(1.4) with $d > 0,$ and $b_{d}$ is the first
value of $r$ such that $w(b_{d}) = \alpha$.
In terms of $y_{\lambda},$ this becomes
$$ 1 \leq y_{\lambda}(r)
  \leq  e^{{M \over 4(m+1)}{r^{2} \over \lambda^{2}}}
         \ {\rm for} \ 0 \leq r \leq \lambda b_{d}. \eqno (4.9)$$
\vskip .2 in
We split the integration interval in (4.7) into the subintervals
$(0, \lambda b_{d})$ and $( \lambda b_{d}, r).$
For the integral over $(0, \lambda b_{d})$, we use (4.9) and the fact that
$$ {F(s) \over s^{2}} \geq -C_{1}$$ for some $C_{1}>0$ and we obtain
$$ \lambda^{- {4 \over p-1} - 2}  \int_{0}^{\lambda b_{d}}
 {F(\lambda^{2 \over p-1} s^{m} y_{\lambda}) \over s^{2m-1}}  \, ds
 \geq -C_{1} \lambda^{- {4 \over p-1} - 2}
 \int_{0}^{\lambda b_{d}}
 {\lambda^{4 \over p-1} s^{2m} y^{2}_{\lambda} \over s^{2m-1}}  \, ds$$
$$ \geq -{C_{1} \over \lambda^{2}}
 \int_{0}^{\lambda b_{d}}
  s e^{{M \over 2(m+1)}{s^{2} \over \lambda^{2}}}  \, ds
  = -{C_{1}(m+1) \over M} [ e^{{M \over 2(m+1)} b_{d}^{2}} - 1].  $$
Now substituting $r = b_{d}$ into (4.8) and recalling that
$d= \lambda^{{2 \over p-1} + m}$ we obtain
$$ \lambda^{{2 \over p-1} + m} b_{d}^{m} \leq w(b_{d}) =\alpha.$$
Hence
$$ b_{d} \to 0 \ {\rm as} \ \lambda \to \infty.$$
Thus, we see that
$$-{C_{1}(m+1) \over M} [ e^{{M \over 2(m+1)} b_{d}^{2}} - 1] \to 0
   \ {\rm as} \ \lambda \to \infty.$$
\vskip .2 in
To estimate the integral over $(\lambda b_{d}, r)$  we recall that
$$ F(s) \geq -C_{2} \ {\rm for \ some} \ C_{2} > 0,$$ so that
$$ \lambda^{- {4 \over p-1} - 2}  \int_{\lambda b_{d}}^{r}
 {F(\lambda^{2 \over p-1} s^{m}y_{\lambda}) \over s^{2m-1}}  \, ds
\geq -C_{2} \lambda^{- {4 \over p-1} - 2}  \int_{\lambda b_{d}}^{r}
 {1 \over s^{2m-1}}  \, ds
 \geq -{ C_{2} \lambda^{- {4 \over p-1} - 2} \over 2m-2}
 {1 \over (\lambda b_{d})^{2m-2}}.      $$
Returning once more to (4.8) and letting $r = b_{d}$ we obtain
$$ {\alpha \over \lambda^{2 \over p-1} e^{{M\over 4(m+1)}b_{d}^{2}}}
    \leq (\lambda b_{d})^{m}.$$
Thus,
$$ -  \lambda^{- {4 \over p-1} - 2} {1 \over (\lambda b_{d})^{2m-2}}
  \geq -{(\lambda b_{d})^{2} \over \lambda^{{4 \over p-1} +2}}
   {\lambda^{4 \over p-1} e^{{M\over 2(m+1)}b_{d}^{2}}  \over \alpha^{2}}
= -{b_{d}^{2} e^{{M \over 2(m+1)}b_{d}^{2}}  \over \alpha^{2}}. $$
As we saw above, $b_{d} \to 0$ as $\lambda \to \infty.$
Thus, this second integral is also bounded below by a constant
which goes to zero as $\lambda \to \infty.$
\vskip .2 in
Thus, we see that
$$ m^{2} \geq E_{\lambda}(r) \geq
  y_{\lambda}^{2} [ {m^{2} \over 2} - C_{1}{r^{2} \over \lambda^{2}}] +
    C_{3}(\lambda)  \ \ {\rm where} \ C_{3}(\lambda) \to 0 \ {\rm as} \
    \lambda \to \infty.$$
Thus, for $r$ in any compact set, if $\lambda$ is chosen large enough we have
$$ y_{\lambda}^{2} \leq M^{2}$$ where $M$ is independent of $\lambda$.

  Next we will show that $  y_{\lambda}'^{2} \leq C.$
Multiplying (4.2) by $r^{2m+1}$ and integrating gives:
$$ -y_{\lambda}' = {\lambda^{{-2 \over p-1}-2} \over r^{2m+1}}
    \int_{0}^{r} s^{m+1} f(\lambda^{2 \over p-1} s^{m} y_{\lambda}) \, ds$$
Since $$ |f(w)| \leq C|w|^{p} + D |w|, $$ for some positive constants
$C$ and $D$, we have
$$| f(\lambda^{2 \over p-1} s^{m} y_{\lambda})|
\leq C \lambda^{2p \over p-1} s^{pm} M^{p}
   + D \lambda^{2 \over p-1} s^{m} M    $$ for sufficiently large $\lambda$.
Thus,
$$ |y_{\lambda}'| \leq
      {A \over r^{2m+1}} \int_{0}^{r} s^{pm +m +1} \, ds
      + {B \over \lambda^{2} r^{2m+1}} \int_{0}^{r} s^{2m+1} \, ds
\leq {A \over pm +m +2} r^{(p-1)m + 1} +
 {B \over 2m+2} { r  \over \lambda^{2}}.$$
Thus, on compact sets we have $$ y_{\lambda}^{2}, y_{\lambda}'^{2} \leq C.  $$
So, by the Arzela-Ascoli Theorem, there
exists a subsequence (again labeled by $\lambda$) such that the $\{ y_{\lambda}
\}$
converge uniformly as $\lambda \to \infty$ to some continuous function $y$.
It remains to show that $y$ satisfies (4.5) - (4.6).

 Since $y_{\lambda}$ is a solution of (4.2) - (4.3) we have
$$ - r^{2m+1}y_{\lambda}' = \int_{0}^{r}
\lambda^{-{2 \over p-1} - 2} s^{m+1} f(\lambda^{2 \over p-1} s^{m} y_{\lambda})
\, ds.  $$
Since $y_{\lambda} \to y$ uniformly and $f$ is of the form (4.4), we see
that the right hand side of the above converges to
$$ j(r) = \int_{0}^{r}  \kappa s^{1+m+pm} |y|^{p-1} y \, ds.  $$
Thus, the sequence $\{ y_{\lambda}'\} $ converges pointwise to the function
${- j(r) \over r^{2m+1}}.$
Since $y$ is continuous, so is $j$. Further, since the $\{ y_{\lambda}'\}$ are
uniformly bounded on compact sets, we can apply the dominated convergence
theorem to
$$ y_{\lambda}(r) = 1 + \int_{0}^{r} y_{\lambda}'(s) \, ds $$
and deduce that
$$ y(r) = 1 - \int_{0}^{r} {j(s) \over s^{2m+1}} \, ds.$$
Therefore, $y'(r) = {- j(r) \over r^{2m+1}}.$
Thus, $$y(0) = 1,$$
$$ y'(0) = \lim_{r \to 0} y'(r) =
\lim_{r \to 0} { - \int_{0}^{r}  \kappa s^{1+m+pm} |y|^{p-1} y \, ds \over
r^{2m+1}}
= \lim_{r \to 0} { - \kappa r^{1+m+pm} |y|^{p-1} y  \over (2m+1) r^{2m}} = 0,$$
and
$$ r^{2m+1}y' = -\int_{0}^{r}  \kappa s^{1+m+mp}|y|^{p-1} y \, ds.  $$
This is equivalent to (4.5) - (4.6).
\vskip .2 in
\noindent
{\bf LEMMA 4.2} : Let $y$ be a solution of (4.5) - (4.6). Then $y$ has at least
one zero.
\vskip .2 in
\noindent
{\bf PROOF OF LEMMA 4.2} : It will be somewhat simpler to show that
$z = r^{m}y$ has at least one zero for $r>0$. We use an argument based on that
of Proposition 3.9 of [5]. The function $z$ satisfies :
$$ z'' + {1 \over r} z' - {m^{2} \over r^{2}} z + \kappa |z|^{p-1}z = 0 $$
and
$$ \lim_{r \to 0} {z \over r^{m}} = 1, \
   \lim_{r \to 0} {z' \over r^{m-1}} = m.$$
Multiplying
by $r^{2} z'$ and integrating on $(0, r)$ gives Pohozaev's Identity :
$$ {1 \over 2} r^{2} z'^{2}  -{m^{2} \over 2} z^{2} +
{\kappa \over p+1}r^{2} |z|^{p+1}
= {2\kappa \over p+1}\int_{0}^{r} s |z|^{p+1} \, ds. \eqno (4.10) $$
\vskip .2 in
Now we assume that $y>0$ for $r>0$ (thus $z > 0$ for $r > 0$) and we will show
that $z \to 0$, $r^{2} |z|^{p+1} \to 0$ as $r \to \infty$ and
$$  \int_{0}^{\infty} s|z|^{p+1} \, ds < \infty.$$
Using (4.10), this will show that $r^{2}z'^{2}$ has a nonzero limit and this
will lead
to a contradiction.
\vskip .2 in
Assuming now that $y > 0,$
multiplying (4.5) by $r^{2m+1}$ and integrating on $(0, r)$ gives:
$$ - r^{2m+1} y' = \int_{0}^{r} \kappa s^{m + mp + 1} y^{p} \, ds.  \eqno
(4.11) $$
Thus, (4.13) implies $ y' < 0.$ Hence, $y$ is decreasing. So,
$$ - r^{2m+1} y' = \int_{0}^{r} \kappa s^{m + mp+1} y^{p} \, ds \geq
y^{p} \int_{0}^{r} \kappa s^{m + mp+1} \, ds
             = {\kappa y^{p} r^{m + mp+2} \over m + mp+2}.   $$
Dividing by $r^{2m+1} y^{p}$ and integrating on $(0,r)$ gives
$$ {y^{-p+1} - 1 \over p-1} \geq {\kappa r^{mp-m+2} \over (m+mp+2)(mp-m+2)}. $$
Hence,
$$ y \leq {C_{m,p} \over r^{m + {2 \over p-1}}}
     \ {\rm where} \ C_{m,p} = [{(m+mp+2)(mp-m+2) \over c(p-1)}]^{1 \over p-1}
.$$
Therefore,
$  z \leq C_{m,p} r^{-{2 \over p-1}}  $
so,
$$ \lim_{r \to \infty} z = 0.   \eqno (4.12) $$
Also,
$ r^{2} z^{p+1} \leq C_{m,p}^{p+1}  r^{-{4 \over p-1}}  $
so,
$$ \lim_{r \to \infty} r^{2} z^{p+1} = 0. \eqno (4.13) $$
Further,
$$ \int_{1}^{\infty} rz^{p+1} \leq
\int_{1}^{\infty} {C_{m,p}^{p+1} \over r^{1 + {4 \over p-1}}}
 = {C_{m,p}^{p+1} (p-1) \over 4}.   \eqno (4.14) $$
Now using (4.12) - (4.14), we can take limits in (4.10) and obtain
$$ \lim_{r \to \infty} {1 \over 2} r^{2} z'^{2}
= {2\kappa \over p+1} \int_{0}^{\infty} r |z|^{p+1} = L < \infty.$$
Clearly, $L \geq 0.$ If $L = 0$ then $\int_{0}^{\infty} r |z|^{p+1} =0$ which
implies $z \equiv 0.$ But this contradicts the fact that $z > 0.$
On the other hand, if $L > 0,$ then
$$ r^{2} z'^{2} \geq L  \ {\rm for \ large} \ r. $$
Thus, $|z'| > 0$ for large $r,$ and since $\lim_{r \to \infty} z = 0$
and $z > 0$ we must then have $z' < 0$ for large $r$.
Therefore, for some $r_{0}$ we have
$$  -z' \geq { \sqrt{L} \over r} \ {\rm for} \ r \geq r_{0}.$$
Hence,
$$ z(r_{0}) - z(r) \geq  \sqrt{L} \log{r \over r_{0}}
    \   {\rm for} \ r \geq r_{0}.    $$
This implies $z(r) \to -\infty$ as $r \to \infty.$
Therefore, we obtain a contradiction to the assumption that $z>0$ for all
$r>0,$ and thus $z,$ and hence $y,$ has at least one positive zero.
This completes the proof of Lemma 4.2.
\vskip .2 in
\noindent
{\bf LEMMA 4.3} : Let $y$ be a solution of (4.5) - (4.6). Then $y$ has
infinitely many zeros.
\vskip .2 in
\noindent
{\bf PROOF OF LEMMA 4.3} : From Lemma 4.2, we know there is a smallest
$r_{1} > 0$ such that $z(r_{1}) = 0$ where $z = r^{m} y.$ By virtue of (4.10)
we know that
$z'(r_{1}) = -A < 0.$ Recall that $$ \tilde E = {1 \over 2}
{z'^{2} \over r^{2m-2}} + {\kappa \over p+1} {  |z|^{p+1} \over r^{2m-2}} +
 {m^{2} \over 2}{z^{2} \over r^{2m}}$$ is a decreasing energy for (4.5).
Thus, $$ \tilde E(r) \leq \tilde E(r_{1}) =
 {1 \over 2} {A^{2} \over r_{1}^{2m-2}}$$ for $r \geq r_{1}.$ In particular,
 for $r \geq r_{1}$ we have,
$$ {|z|^{p+1} \over r^{2m-2}} \leq {p+1 \over 2\kappa}{ A^{2} \over
r_{1}^{2m-2}}.$$
Thus $y$ satisfies (4.5) and :
$$ y(r_{1}) = 0, \ y'(r_{1}) = -{A \over r_{1}^{m}}.$$
The bound on $z$ for $r \geq r_{1}$ yields a corresponding estimate for $y$ :
$$ |y|^{p+1} \leq  { C \over r^{m(p-1) + 2}}. $$
Thus, $y \to 0$ as $r \to \infty,$ and so it follows that $y$ must have
a local minimum, $y_{1}$, at $r = t_{1} > r_{1}.$ So $y$ satisfies the
initial value problem consisting of (4.5) subject to
$$ y(t_{1}) = y_{1}, \ y'(t_{1}) = 0.$$
We may now use the same argument as in Lemma 4.2, replacing $r=0$ with
$r=t_{1}$ to show that $y$ has a second zero at $ r = r_{2} > r_{1}$.
Proceeding inductively,
we can show that $y$ has infinitely many zeros.
\vskip .2 in
\noindent
{\bf LEMMA 4.4} : Denote by $w(r,d)$ the solution to (1.3) - (1.4).
Let $d_{0}$ be a value for which $w(r,d_{0})$ has exactly
$k$ zeros and $w(r,d_{0}) \to 0$ as $r \to \infty$. If $|d-d_{0}|$ is
sufficiently small, then $w(r,d)$ has at most $k+1$ zeros on $[0, \infty).$
\vskip .2 in
\noindent
{\bf PROOF OF LEMMA 4.4} :
The proof is similar to those for Lemmas 3.1 - 3.5.
We wish to show that for $d$ near $d_{0}$, $w(\cdot, d)$ has at most
$(k+1)$ zeros in $[0, \infty).$ So we suppose there is a sequence of values
$d_{j}$
converging to $d_{0}$ such that $w(\cdot, d_{j})$ has at least $(k+1)$ zeros
in $[0, \infty)$. (If there is no such sequence, then the lemma is proven.)
We write $w_{j}(r) \equiv w(r, d_{j})$ and we denote by $r_{j}$ the $(k+1)$st
zero of $w_{j}$, counting from the smallest.

   Because of the continuous dependence of the solution on initial conditions,
we know that $w_{j} \to w_{0} \equiv w(\cdot, d_{0})$ and $w_{j}' \to w_{0}'$
uniformly on compact sets as $j \to \infty$. In particular, let $[0,L)$
be any bounded interval containing the $k$ zeros of $w_{0}$. Then,
for sufficiently large $j$, the function $w_{j}$ has exactly $k$ zeros in $[0,
L).$
Thus, $r_{j} \to \infty$ as $j \to \infty$.

   Let $b_{j}$ be the smallest number greater than $r_{j}$ such that
$|w_{j}(b_{j})| = \alpha$. The existence of $b_{j}$ is guaranteed by Lemma 3.1.
Let $a_{j}$ be the smallest number greater than $r_{j}$ such that
$|w_{j}(a_{j})| = { \alpha \over 2}.$

   Let $q_{j}$ be the largest number less than $r_{j}$ such that
$|w_{j}(q_{j})| = \gamma$. That there is such a number $q_{j}$ can be seen as
follows. Let $p_{j}$ be the location of a local extremum between the $k$th and
$(k+1)$st zeros of $w_{j}$. Evaluating Pohozaev's Identity between $p_{j}$ and
$r_{j}$ gives

$$ 0 < {1 \over 2} r_{j}^{2}w_{j}'^{2}(r_{j}) + {m^{2} \over 2}w_{j}^{2}(p_{j})
=
  p_{j}^{2}F(w(p_{j})) + \int_{p_{j}}^{r_{j}} s F(w(s)) \, ds. $$
It follows that $F(w(r)) > 0$ for some $r$ in $(p_{j}, r_{j})$, and hence
$|w(r)| > \gamma$ for some $r$ in $(p_{j}, r_{j})$. Thus, there is a largest
number $q_{j}$ less than $r_{j}$ such that $|w_{j}(q_{j})| = \gamma.$

   As in Lemma 3.3, we may now verify the following.

\noindent
{\bf CLAIM} : $b_{j} - a_{j} \geq K^{2} > 0$, where $K$ is a constant
independent of $j$ for $j$ sufficiently large.

\noindent
{\bf PROOF OF CLAIM} : Evaluating Pohozaev's Identity between $q_{j}$ and $r$
gives

$$ r^{2}[{1 \over 2} w_{j}'^{2}(r) + F(w_{j}(r))]  =
  {m^{2} \over 2}(w_{j}^{2}(r) - \gamma^{2}) +
  {1 \over 2}q_{j}^{2} w_{j}'^{2}(q_{j}) +
   2 \int_{q_{j}}^{r} s F(w_{j}(s)) \, ds.  $$

Using the facts that $|w_{j}(r)| \leq \gamma$ and $F(w_{j}(r)) \leq 0$ for
$q_{j} \leq r \leq b_{j},$ we obtain
$$ {1 \over 2} w_{j}'^{2}(r) + F(w_{j}(r)) \leq {1 \over 2} w_{j}'^{2}(q_{j})$$

  Now, $|w_{j}(q_{j})| = \gamma,$ and $w_{j}$ tends to $w_{0}$ uniformly
on compact sets as $j \to \infty$. Since $w_{0}(r) \to 0$ as $r \to \infty$,
it follows that $q_{j}$ has the finite limit $q_{0}$ as $ j \to \infty$,
where $|w_{0}(q_{0})| = \gamma.$ Hence, $w_{j}'(q_{j}) \to w_{0}'(q_{0})$
as $j \to \infty$, so that
$$  {1 \over 2} w_{j}'^{2}(r) + F(w_{j}(r)) \leq D \eqno (4.15) $$ for
$q_{j} \leq r \leq b_{j},$ where $D > 0$ is a constant that is independent of
$j$ for sufficiently large $j$.

   Now Lemma 3.1 shows that $w_{j}'(r) \neq 0$ for $r$ in $(r_{j}, b_{j}]$
so from (4.15) we have
$$ 0 < \int_{\alpha \over 2}^{\alpha} {dy \over \sqrt{D - F(y)} }
= \int_{a_{j}}^{b_{j}} { |w_{j}'(r)| \, dr \over \sqrt{ D - F(w_{j}(r))} }
\leq \int_{a_{j}}^{b_{j}} \sqrt{2} \, dr = \sqrt{2}(b_{j} - a_{j})$$
for sufficiently large $j$. This proves the claim, with
$K^{2} \equiv{1 \over 2 \sqrt{2}}
  \int_{\alpha \over 2}^{\alpha} {dy \over \sqrt{D - F(y)} }$

  As in Lemma 3.4, we may verify the following.

\noindent
{\bf CLAIM} :  For sufficiently large $j$, $|w_{j}(r)| < \gamma$ for all
$r > r_{j}.$

\noindent
{\bf PROOF OF CLAIM} : Suppose on the contrary that there is a smallest
$c_{j} > r_{j}$ such that $|w_{j}(c_{j})| = \gamma.$ Evaluating Pohozaev's
Identity between $q_{j}$ and $c_{j}$ gives
$$ {1 \over 2} c_{j}^{2} w_{j}'^{2}(c_{j})   =
  {1 \over 2}q_{j}^{2} w_{j}'^{2}(q_{j}) +
   2 \int_{q_{j}}^{c_{j}} s F(w_{j}(s)) \, ds.  \eqno (4.16) $$
Since $F(w(s)) \leq 0$ on $(c_{j}, q_{j})$, and $F$ is decreasing on
$[{\alpha \over 2}, \alpha]$, as in Lemma 3.4 we have
$$ \int_{q_{j}}^{c_{j}} sF(w_{j}(s)) \, ds
  \leq \int_{a_{j}}^{b_{j}} sF(w_{j}(s)) \, ds
  \leq {1 \over 2} F({\alpha \over 2})  (b_{j}^{2} - a_{j}^{2})
  \leq {K^{2} \over 2} F( {\alpha \over 2})    (b_{j} + a_{j}). $$

Now $q_{j}^{2} w_{j}'^{2}(q_{j}) \to q_{0}^{2} w_{0}'^{2}(q_{0})$ and
$a_{j} + b_{j} \to \infty$ as $j \to \infty$. Thus, the right-hand side of
(4.16) tends to $-\infty$ as $j \to \infty$, whereas the left-hand side is
positive. This contradicts the assumption that there is $c_{j} > r_{j}$ with
$|w_{j}(c_{j})| = \gamma,$ and proves the claim.

  Finally, to complete the proof of Lemma 4.4, suppose that $w_{j}$ has another
zero $t_{j} > r_{j}.$ Then there is a local extremum for $w_{j}$ at a location
$s_{j}$ with $r_{j} < s_{j} < t_{j}.$ Evaluating Pohozaev's Identity between
$s_{j}$ and $t_{j}$ gives
$$ {1 \over 2} t_{j}^{2} w_{j}'^{2}(t_{j}) +
  {m^{2} \over 2}w_{j}^{2}(s_{j}) =
  s_{j}^{2} F(w_{j}(s_{j})) +
   2 \int_{s_{j}}^{t_{j}} s F(w_{j}(s)) \, ds.  $$

This implies that $F(w_{j}(r)) > 0$ for some $r$ between $s_{j}$ and $t_{j}$.
But for sufficiently large $j$, $|w_{j}(r)| < \gamma$ for all $r > r_{j}$,
hence $F(w_{j}(r)) < 0$ for $r$ between $s_{j}$ and $t_{j}$. This contradiction
shows that for $j$ sufficiently large, there is no zero of $w_{j}$ larger
than $r_{j}$. This completes the proof of Lemma 4.4

\parindent = 20 pt
\hsize = 6.5 in
\vsize = 8.5 in
\voffset = 0 in
\hoffset = 0 in
\vskip .4 in
\noindent
\centerline{5. PROOF OF THE MAIN THEOREM}
\vskip .2 in
\noindent
{\bf PROOF OF MAIN THEOREM} : We define
$$A_{0} = \{ d > 0 | w(r, d) \ {\rm has \ no \ positive \ zeros} \}.   \eqno
(5.1) $$
Lemma 3.4 shows that the set $A_{0}$ is nonempty. Also,
Lemmas 4.1 - 4.3 imply that the set $A_{0}$ is bounded above. Let
$ d_{0} = \sup A_{0}. $
We will show that
$$ w(r, d_{0}) > 0, \ {\rm and} \ w(r, d_{0}) \to 0 \ {\rm as} \ r \to
\infty.$$
First, if $w(r, d_{0})$ has a zero at some finite $r$, then continuity of
$w(r,d)$ on $d$ implies that $w(r,d)$ has a zero for $d$ slightly smaller than
$d_{0}$. This contradicts the definition of $d_{0}$. Thus,
$$ w(r, d_{0}) > 0 {\rm \ for \ all} \ r > 0.\eqno (5.3)$$
Next, either $$ w \ {\rm is \ monotone \ for \ large} \ r $$ or
$$ w(r, d_{0}) {\rm \ has \ positive \ local \ minima \ at \ arbitrarily
    \ large \  values \ of} \ r.  $$
We will show this second case is impossible.
\vskip .1 in
\noindent
{\bf CLAIM } : If $w$ has positive local minima at $M_{k}$ where
$M_{k} \to \infty$, then $\limsup_{k \to \infty}w(M_{k}) \leq \beta.$
\vskip .1 in
\noindent
{\bf PROOF OF CLAIM } :
At a minimum $M_{k}$ of $w$, we have $w'(M_{k}) = 0 $ and
$w''(M_{k}) \geq 0.$ Substituting into (1.3) gives
$$  -{m^{2} \over M_{k}^{2}} w(M_{k}) + f(w(M_{k})) \leq 0.$$
Thus, since $w(M_{k}) > 0$, we obtain
$$ { f(w(M_{k})) \over w(M_{k}) } \leq {m^{2} \over M_{k}^{2}}. $$
Thus, since $M_{k} \to \infty$ as $k \to \infty$ we have
$$ \limsup_{k \to \infty} { f(w(M_{k})) \over w(M_{k}) } \leq 0.$$
This implies
$$\limsup_{k \to \infty} w(M_{k}) \leq \beta.$$
This completes the proof of CLAIM.
\vskip .2 in

  Now we will show that it is impossible for $w(r,d_{0})$ to have positive
local
minima at arbitrarily large values of $r$. From the above CLAIM,
for sufficiently large $k$ we have
$$   w(M_{k},d_{0}) \leq {3 \over 4}\beta + {1 \over 4}\gamma. $$
Now, we choose $d$ slightly larger than $d_{0}$ so that $w(r,d)$ has a zero at
$z$. By virtue of continuous dependence on initial values,
for $d$ close enough to $d_{0}$, $w(r,d)$ will also have positive local
minima at $N_{k}$ for $k = 1, \cdots, K,$ where
$N_{K}$ is the largest value of $r$ less than $z$ at which $w$ has a minimum.
Evaluating Pohozaev's Identity between $0$ and $z$ gives
$$ 0 \leq {1 \over 2}z^{2} w'^{2}(z) = 2 \int_{0}^{z}rF(w) \, dr. \eqno (5.6)
$$
Also, evaluating between $0$ and $N_{K}$ gives
$$ N_{K}^{2}F(w(N_{K}))  = {m^{2} \over 2}w^{2}(N_{K})
+ 2 \int_{0}^{N_{K}}rF(w) \, dr. \eqno (5.7)$$
We may furthermore choose $d$ sufficiently close to $d_{0}$ so that
$$ w(N_{K}, d) \leq {1 \over 2}\beta + {1 \over 2}\gamma < \gamma. \eqno
(5.8)$$
Thus, $$ F(w(N_{K})) \leq 0.$$ So we see from (5.7) that
$$ 2 \int_{0}^{N_{K}}rF(w) \, dr \leq 0.  \eqno (5.9) $$
Combining (5.6) and (5.9) we have
$$ 2 \int_{N_{K}}^{z}rF(w) \, dr \geq 0.   \eqno (5.10)$$
Thus, there exists a $c$ with
$$ N_{K} < c < z  {\rm \ such \ that} \ w(c) = \gamma.$$
Further, $c$ may be chosen so that $w' \geq 0$ on $[N_{K}, c]$
because $N_{K}$ is the largest value of $r$ at which $w$ has a minimum before
the zero, $z,$ of $w$. Pohozaev's Identity evaluated between $N_{K}$ and $c$
gives
$$ 0 \leq {1 \over 2}c^{2} w'^{2}(c) =   N_{K}^{2}F(w(N_{K})) +
 {m^{2} \over 2}(\gamma^{2} - w^{2}(N_{K})) + 2 \int_{N_{K}}^{c}rF(w) \, dr
 \eqno (5.12) $$
$$\leq {m^{2} \over 2}(\gamma^{2} - w^{2}(N_{K})) +
   2 \int_{N_{K}}^{c}rF(w) \, dr.   $$
We will now show that the right hand side of (5.12) is negative for $d$
close enough to $d_{0}$ and thus obtain a contradiction to the assumption
that $w(r,d_{0})$ has local minima at arbitrarily large values of $r$.
\vskip .2 in

  We will estimate the integral on the right.
Let $a, b$ be such that $ N_{K} < a < b < c$ and
$w(a) = {1 \over 2}\beta + {1 \over 2}\gamma, \ w(b)
= {1 \over 4}\beta + {3 \over 4} \gamma.$
Then since $w' \geq 0,$ and $F(w) \leq 0$ on $[a,b]$,  and $f(w)= F'(w) \geq 0$
on
$[ {\beta + \gamma \over 2}, {\beta + 3\gamma \over 4}]$
we obtain :
$$ 2 \int_{N_{K}}^{c}rF(w) \, dr \leq  2 \int_{a}^{b}rF(w) \, dr
\leq F({\beta + 3\gamma \over 4}) (b^{2} - a^{2}).$$
Thus, from (5.12) we see that
$$ 0 \leq {m^{2} \over 2}(\gamma^{2} - w^{2}(N_{K}))
 + F({\beta + 3\gamma \over 4}) (b^{2} - a^{2}). \eqno (5.13) $$
\vskip .2 in

  We will next show that $b-a \geq \epsilon > 0$ where $\epsilon$ is
independent
of $d$. To show this, we note that for $N_{K} \leq r \leq c$ we have
$$ \int_{0}^{r} 2 sF(w) \, ds = \int_{0}^{N_{K}} 2 sF(w) \, ds
+ \int_{N_{K}}^{r} 2 sF(w) \, ds.$$
{}From (5.9) we see that the first integral on the right is negative.
Also, the second integral is negative because
$0 < w \leq \gamma$ on $[N_{K}, c]$ and so $F(w) \leq 0.$
Thus, Pohozaev's Identity between $0$ and $r$ gives
$$ {1 \over 2} w'^{2} + F(w) \leq  {m^{2} \over 2}{w^{2} \over r^{2}}
 \ {\rm for} \ N_{K} \leq r \leq c.$$
Choosing $d$ close enough to $d_{0}$ we can always ensure that
$N_{K} \geq 1.$
Thus, $$ {1 \over 2} w'^{2} + F(w) \leq C \equiv {m^{2} \over 2}
\gamma^{2} \ {\rm for} \ N_{K} \leq r \leq c. $$
Since $w' \geq 0$ on $[N_{K}, c]$, we obtain
$$  {w' \over \sqrt{C - F(w)}} \leq \sqrt{2}
    \ {\rm for} \ N_{K} \leq r \leq c. $$
Now since $[a,b] \subset [N_{K}, c]$, we can integrate on $[a,b]$ and get
$$ \int_{\beta + \gamma \over 2}^{\beta + 3\gamma \over 4}
 {ds \over \sqrt{C - F(s)}} =
\int_{a}^{b} {w' \over \sqrt{C - F(w)}} \, dr
\leq \int_{a}^{b} \sqrt{2} \, dr = \sqrt{2} (b-a).  $$
We have thus established that $b-a \geq
    {1 \over \sqrt{2}} \int_{\beta + \gamma \over 2}^{\beta + 3\gamma \over 4}
 {ds \over \sqrt{C - F(s)}} =\epsilon > 0,$ as claimed.

  Now, since $b^{2} - a^{2} = (b-a)(b+a)$ and
$ b+a > 2N_{K} \to \infty $ as $d \to d_{0}$
we obtain that $b^{2} - a^{2} \to \infty$ as $d \to d_{0}.$ Therefore, the
right hand side
of (5.13) approaches $ -\infty$ as $d \to d_{0}$. This is the required
contradiction
and hence $w(r, d_{0})$ must be monotone for large values of $r$.

 Next, an argument similar to that in the proof of Lemma 3.2 shows that the
monotonicity of $w(r,d_{0})$ for large $r$ implies that $w(r,d_{0})$ is
bounded in $r$. We thus have
$$ \lim_{r \to \infty}w(r, d_{0}) = \zeta \geq 0. \eqno (5.14)$$
Taking limits in
$$ {1 \over 2} w'^{2} + F(w) = {m^{2} \over 2}{w^{2} \over r^{2}}
   + {2 \over r^{2}}\int_{0}^{r} sF(w) \, ds $$
we obtain
$$ \lim_{r \to \infty}{1 \over 2} w'^{2} + F(\zeta) = F(\zeta).  $$
(When taking the limit of the right-hand side we use the following fact :
if $g$ is continuous and $\lim_{r \to \infty} g(r) = g_{0}$, then
$ \lim_{r \to \infty} {2 \over r^{2}}\int_{0}^{r} sg(s) \, ds = g_{0}$.
Proof : $|{2 \over r^{2}}\int_{0}^{r} s[g(s) - g_{0}] \, ds| \leq
   {2 \over r^{2}}\int_{0}^{R} s|g(s) - g_{0}| \, ds +
   {2 \over r^{2}}\int_{R}^{r} s|g(s) - g_{0}| \, ds
   \leq { 2MR^{2} \over r^{2}} +  {2 \over r^{2}}\int_{R}^{r} s \epsilon \, ds
   =  { 2MR^{2} \over r^{2}}  + \epsilon.) $
Thus, $$ \lim_{r \to \infty} w'(r, d_{0}) = 0. \eqno (5.15)$$
Also, from (1.3) it follows that
$$ \lim_{r \to \infty} w'' = -f(\zeta).$$
So, we must have $$ f(\zeta) = 0.   \eqno (5.16)$$
We finally need to show that $\zeta = 0.$
If not, that is if $\zeta > 0,$ then
since $w(r,d) \to w(r,d_{0})$ uniformly on compact sets,
given $\epsilon > 0$ we can choose a large value, $q,$ and $d > d_{0}$
sufficiently close to $d_{0}$ such that
$$ \zeta - \epsilon \leq  w(q,d) \leq \zeta + \epsilon.$$
Now, we know $w(r,d)$ has a zero $z > q$. Evaluating Pohozaev's Identity
between $q$ and $z$ gives
$$ 0 \leq {1 \over 2}z^{2}w'^{2}(z) =
q^{2}[{1 \over 2}w'^{2}(q) - {m^{2} \over 2}{w^{2}(q) \over q^{2}} + F(w(q))] +
2 \int_{q}^{z}rF(w) \, dr.$$
For $q$ sufficiently large, and for
$d$ close enough to $d_{0}$ but slightly larger than $d_{0}$,
the term in brackets is close to $F(\zeta) < 0$
because $w(r,d) \to w(r,d_{0})$ uniformly on compact sets and
$w(r,d_{0}) \to \zeta$ and $w'(r,d_{0}) \to 0$ as $r \to \infty$.
Thus, $$  \int_{q}^{z}rF(w)  \geq 0.$$
So, there is an $s$ with $q < s < z$ such that $w(s) = \gamma$
and $\beta < w(r) < \gamma$ on $(q,s).$
Evaluating Pohozaev's Identity between $q$ and $s,$ and noting that
$F(w) \leq 0$  for $r \in [q,s],$ we obtain :
$$ 0 \leq {1 \over 2}s^{2}w'^{2}(s)
  \leq  q^{2}[{1 \over 2}w'^{2}(q) + {m^{2} \over 2}{\gamma^{2} \over q^{2}}
     - {m^{2} \over 2}{w^{2}(q) \over q^{2}} + F(w(q))]  .$$
As above, for $d$ close enough to $d_{0}$ and for large enough $q$
the expression in brackets is negative, and this is a contradiction.
Thus, we must have $$\zeta = \lim_{r \to \infty} w(r, d_{0}) = 0.$$
This completes the proof of the existence of a positive solution of
(1.3)-(1.4) with $w(r) \to 0$ as $r \to \infty.$
\vskip .5 in
\noindent
Next, we define
$$A_{1} = \{ d > d_{0} |w(r,d) {\rm \ has \ at \ most \ one \ positive \ zero.}
\} $$
It follows from the definition of $d_{0}$ and Lemma 4.4 that $A_{1}$ is
nonempty. From Lemmas 4.2 - 4.3 it follows that $A_{1}$ is bounded
above. Thus, we define
$$ d_{1} = \sup A_{1}. $$ As above we can show that
$$ w(r,d_{1}) {\rm \ has \ exactly \ one \ zero, \ and \ }
   w(r,d_{1}) \to 0 {\rm \ as \ } r \to \infty.$$
\vskip .2 in
 Proceeding inductively, we can find solutions that tend to zero at infinity
and with any prescribed number of zeros. This completes the proof
of the MAIN THEOREM.

\vskip .4 in
\centerline{6. EXPONENTIAL DECAY}
\vskip .2 in

  Because $\lim_{s \to 0} {f(s) \over s} = -\sigma^{2}$, we expect that
solutions $w$ of (1.3) that have $\lim_{r \to \infty} w(r) = 0$ will be
governed
for large $r$ by the linearized equation
$$ w'' + {1 \over r}w' - {m^{2} \over r^{2}} w - \sigma^{2} w = 0. \eqno
(6.1)$$
If $\sigma > 0$, equation (6.1) is the modified Bessel equation, whose
decaying solution $K_{m}(\sigma r)$ has the asymptotic behavior
$$ K_{m}(\sigma r) = \sqrt{\pi \over 2 \sigma} {1 \over \sqrt{r}}
           e^{-\sigma r}( 1 + O({1 \over r})) \ {\rm as} \ r \to \infty.$$
Here we content ourselves with showing that
$$ w(r) = O(e^{-\rho r}) \ {\rm as} \ r \to \infty \ {\rm for \ all} \ \rho
        \in (0, \sigma). $$

   To do so, let $\sigma > 0$ and consider a solution to (1.3) - (1.4) such
that $\lim_{r \to \infty} w(r) = 0.$ The proof of the main theorem shows that
there is an $r_{0} > 0$ so large that $w$ is monotonic for $r> r_{0}.$ Without
loss of generality we assume $w(r) >0$ and $w'(r) < 0$ for all $r > r_{0}$.
Then from (1.3) we have
$$ w'' + f(w) = {m^{2} \over r^{2}} w - {1 \over r} w' > 0
 \ {\rm for \ all} \ r > r_{0}.$$
Because $\lim_{w \to 0} {f(w) \over w} = -\sigma^{2}$, for every
$\rho$ with $0 < \rho < \sigma$, there is $\epsilon_{\rho} > 0$ such that
$f(w) < -\rho^{2} w$ for all $w$ in $(0, \epsilon_{\rho}).$ Choose $r_{\rho} >
r_{0}$
so large that $0 < w(r) < \epsilon_{\rho}$  for all $r \geq r_{\rho}$. Then
$f(w(r)) < -\rho^{2} w(r)$ for all $r\geq r_{\rho}$, so we have
$$ w'' > -f(w) > \rho^{2} w \ {\rm for \ all} \ r \geq r_{\rho}. $$ Multiplying
this inequality by $w'(r) < 0$ gives
$$ (w'^{2})' < \rho^{2} (w^{2})'.$$ Because $w(r) \to 0$ and $w'(r) \to 0$ as
$r \to \infty$, we may integrate each side of the inequality from $r$ to
$\infty$
to obtain
$$ w'^{2} < \rho^{2} w^{2} \ {\rm for} \ r \geq r_{\rho}.$$ Using the facts
that $w> 0$ and $w'<0$ for $r\geq r_{\rho}$, we take square roots, divide by
$w$, and integrate from $r_{\rho}$ to $r$ to obtain
$$ w(r) < w(r_{\rho}) e^{ - \rho(r-r_{\rho})} {\rm \ for} \ r \geq r_{\rho},$$
which establishes that
$w(r) = O(e^{-\rho r}) \ {\rm as} \ r \to \infty.$

\vskip .4 in
\centerline{REFERENCES}
\vskip .2 in
\hangindent .3 in
\noindent
[1] \hskip .1 in
H. Berestycki and P. L. Lions, {\it Nonlinear scalar field equations,
I and II.}  Arch. Rat. Mech. Anal. 82  (1983) 313 - 375.
\vskip .1 in
\hangindent .3 in
\noindent
[2] \hskip .1 in
M. Berger, {\it Nonlinearity and Functional Analysis.}  Academic
Press, New York (1977).
\vskip .1 in
\hangindent .3 in
\noindent
[3] \hskip .1 in
E. Deumens and H. Warchall, {\it Explicit construction of all spherically
symmetric solitary waves for a nonlinear wave equation in multiple
dimensions.}  Nonlin. Anal. T. M. A. 12 (1988) 419 - 447.
\vskip .1 in
\hangindent .3 in
\noindent
[4] \hskip .1 in
B. Gidas, W-M. Ni, L. Nirenberg, {\it Symmetry and related properties
via the maximum principle.} Comm. Math. Phys. 68 (1979) 209-243.
\vskip .1 in
\hangindent .3 in
\noindent
[5] \hskip .1 in
A. Haraux and F. B. Weissler, {\it Non-uniqueness for a semilinear
initial value problem.} Indiana Univ. Math. J, 31 (1982), 167 - 189.
\vskip .1 in
\hangindent .3 in
\noindent
[6] \hskip .1 in
C. Jones and T. K\"upper, {\it On the infinitely many solutions of a
semilinear elliptic equation.} SIAM J. Math. Anal. 99 (1987), 803-835.
\vskip .1 in
\hangindent .3 in
\noindent
[7] \hskip .1 in
G. King, {\it Explicit multidimensional solitary waves.} University
of North Texas Master's Thesis, August 1990.
\vskip .1 in
\hangindent .3 in
\noindent
[8] \hskip .1 in
P. L. Lions, {\it Solutions complexes d'\'equations elliptiques
semilin\'eaires dans $R^{N}$.} C. R. Acad. Sc. Paris 302(\#19)(1986)673-676.
\vskip .1 in
\hangindent .3 in
\noindent
[9] \hskip .1 in
K. McLeod and J. Serrin, {\it Uniqueness of positive radial solutions of
$\Delta u + f(u) = 0$ in $R^{N}$.} Arch. Rat. Mech. Anal. 99 (1987) 115 - 145.
\vskip .1 in
\hangindent .4 in
\noindent
[10] \hskip .1 in
K. McLeod, W.C. Troy, and F.B. Weissler,
{\it Radial solutions of $\Delta u + f(u) = 0$ with
prescribed number of zeros.} J. Diff. Eq.,  83 (1990) 368 - 378.
\vskip .1 in
\hangindent .4 in
\noindent
[11] \hskip .1 in
L. A. Peletier and J. Serrin, {\it Uniqueness of positive solutions of
semilinear equations in $R^{n}$.} Arch. Rat. Mech. Anal. 81 (1983)
 181-197.
\vskip .1 in
\hangindent .4 in
\noindent
[12] \hskip .1 in
W. Strauss, {\it Existence of solitary waves in higher dimensions.}
Comm. Math. Phys. 55 (1977) 149 - 162.
\bye